\documentclass[twocolumn,amsmath,amssymb]{revtex4}
\usepackage{graphicx}

\newcommand{\aj}{AJ}
\newcommand{\apjs}{ApJS}
\newcommand{\apjl}{ApJL}
\newcommand{\mnras}{MNRAS}

\newcommand{\aap}{A\&A}

\newcommand{\apss}{Ap\&SS}

\newcommand{\na}{New Astronomy}

\newcommand{\ep}{\epsilon}

\begin{document}

\title{Cusps in the center of galaxies: a real conflict with observations
or a numerical artefact of cosmological simulations?}
\author{A. N. Baushev}
 \affiliation{Bogoliubov Laboratory of Theoretical Physics, Joint Institute for Nuclear Research,
141980 Dubna, Moscow region, Russia}
 \affiliation{Departamento de Astronom\'ia, Universidad de
Chile, Casilla 36-D, Correo Central, Santiago, Chile}
\author{L.~del~Valle}
 \affiliation{Departamento de Astronom\'ia, Universidad de
Chile, Casilla 36-D, Correo Central, Santiago, Chile}
\author{L.E.~Campusano}
 \affiliation{Departamento de Astronom\'ia, Universidad de
Chile, Casilla 36-D, Correo Central, Santiago, Chile}
\author{A.~Escala}
 \affiliation{Departamento de Astronom\'ia, Universidad de
Chile, Casilla 36-D, Correo Central, Santiago, Chile}
\author{R.R.~Mu\~noz}
 \affiliation{Departamento de Astronom\'ia, Universidad de
Chile, Casilla 36-D, Correo Central, Santiago, Chile}
\author{G.A.~Palma}
 \affiliation{Departamento de F\'isica, FCFM, Universidad de Chile, Blanco Encalada 2008, Santiago, Chile}

\date{\today}

\begin{abstract}
Galaxy observations and N-body cosmological simulations produce conflicting dark matter halo
density profiles for galaxy central regions. While simulations suggest a cuspy and universal
density profile (UDP) of this region, the majority of observations favor variable profiles with a
core in the center. In this paper, we investigate the convergency of standard N-body simulations,
especially in the cusp region, following the approach proposed by \protect{\citep{13}}. We simulate
the well known Hernquist model using the SPH code Gadget-3 and consider the full array of dynamical
parameters of the particles.  We find that, although the cuspy profile is stable, all integrals of
motion characterizing individual particles suffer strong unphysical variations along the whole
halo, revealing an effective interaction between the test bodies. This result casts doubts on the
reliability of the velocity distribution function obtained in the simulations. Moreover, we find
unphysical Fokker-Planck streams of particles in the cusp region. The same streams should appear in
cosmological N-body simulations, being strong enough to change the shape of the cusp or even to
create it. Our analysis, based on the Hernquist model and the standard SPH code, strongly suggests
that the UDPs generally found by the cosmological N-body simulations may be a consequence of
numerical effects. A much better understanding of the N-body simulation convergency is necessary
before a 'core-cusp problem' can properly be used to question the validity of the CDM model.
\end{abstract}

\maketitle

\section{Introduction}
Results of N-body simulations come into increasing conflict with observations of the dark matter
(DM) distribution in the central regions of dwarf galaxies. Astronomical observations favor
relatively soft cored density profiles \citep{mamon2011, deblok2001, bosma2002, oh2011,
governato2012, tollerud2012, delpopolo2016}. On the contrary, N-body simulations of cold dark
matter tell us that dark matter halos have a universal shape, independent of the halo mass and
initial density fluctuation spectrum, and that the central universal density profile (hereafter
UDP) is cuspy. The first works on the subject proposed the Navarro-Frenk-White profile (hereafter
NFW) that behaves as $\rho\propto r^{-1}$ at the center. Later simulations \citep{mo09,navarro2010}
favor the Einasto profile with a finite central density. However, the obtained Einasto index is so
high (typically $n\sim 5-6$) that the profile is still cuspy and close to the NFW one.

For a time, there was a hope that the 'core-cusp problem' would disappear once the baryon
contribution is taken into account. However, recent simulations including baryon matter have rather
amplified the problem \citep{diversity}: apart from the profile disagreement, a more fundamental
difficulty was found. Of course, the presence of baryons in simulations changes the density
profile, but it remains almost universal for all the halos, while the profiles of real galaxies are
extremely varied. The conflict between simulations and observations might suggest that the cold DM
paradigm is wrong. However, before reaching this conclusion,  the accuracy and convergence of the
simulations should be scrutinized. For instance, the overestimation of the energy exchange between
the test bodies that may occur in the N-body simulations leads to the cusp formation \citep{15}. If
the energy evolution during the halo formation is limited, then the density profile of the formed
halo resembles more closely the observed one \citep{16}.

As an example, the overestimation of the particle energy exchange may be due to the unphysical pair
collisions of the test bodies. Its importance may be characterized by the relaxation time
\citep[eqn. 1.32]{bt}
\begin{equation}
\tau_r =\dfrac{N(r)}{8\ln\Lambda}\cdot\tau_d \label{relaxation_time}
\end{equation}
where $N(r)$  is the number of test bodies inside a sphere of radius $r$, $\ln\Lambda$ is the
Coulomb logarithm, $\tau_d=(6\pi G\bar\rho(r))^{-1/2}$ is the characteristic dynamical time of the
system at radius $r$, $\bar\rho(r)$ is the average density inside $r$.
Equation~\ref{relaxation_time} has two important consequences. First, $\tau_r$ depends on the
smoothing radius of the N-body simulations only through $\Lambda$, i.e., only logarithmically
\citep{13}. Therefore, the influence of the unphysical collisional relaxation cannot be decreased
much by the smoothing of the test body potentials. Second, since the number of dark matter
particles is huge ($\sim 10^{60}$, if dark matter consists of elementary particles), the
collisional relaxation plays no role in nature, being a purely numerical effect.

The algorithm stability is the critical point of N-body simulations: the Miller's instability makes
the Liapunov time comparable with the dynamical time of the system \citep{miller1964}. Even if we
take into account the specificity of N-body algorithms, like the potential smoothing, the
instability development time is much shorter than $\tau_r$ and remains comparable with the
dynamical time at the given radius $\tau_d(r)$ \citep{valluri2000,hut2002}. However, different
N-body codes, with various versions of the Poisson solvers, integration algorithms etc., lead to
final halos with the above-mentioned UDP, which is almost the same and close to NFW. Therefore, it
is widely believed that the universal profile is physically meaningful and that it describes real
halos, even though the orbits of individual test bodies have no physical significance
\citep[section 4.7.1(b)]{bt}. The aim of this paper is to question this opinion.

Indeed, the convergency criteria of N-body simulations used at present are exclusively based on the
density profile stability. \citep{power2003} found that the cusp of the UDP remains stable at least
until $t=1.7\tau_r$ and then a core forms. On this basis \citep{power2003} supposed that the core
formation is the first sign of the collision influence and offered the most extensively used
criterion for simulation convergency $t<1.7\tau_r$. The acceptance that the collisions have no
effect even if the simulation time exceeds $\tau_r$ seems surprising. However, later convergency
tests (also based only on the stability of the density profile) suggested even softer criteria
\citep{hayashi2003,klypin2013}. In this paper we perform a more sophisticated convergency test,
going beyond the density profile analysis and considering the full array of the dynamical
parameters of the particles.

\section{Calculations}
\subsection{The main idea}
In order to test the N-body convergency, we follow the method offered in \citep{13}. We simulate
the well-known Hernquist model with the density profile $\rho(r)=Ma/[2\pi r(r+a)^3]$ (where $a$ is
the scale radius and $M$ is the total halo mass), and with the isotropic velocity distribution at
each point \citep{hernquist1990}. The model is spherically symmetric and fully stable, i.e., the
density and velocity profiles should not change with time. We chose the Hernquist model because it
is close to the NFW and behaves exactly as the NFW ($\rho\propto r^{-1}$) in the central region,
but it has a known analytical solution for the stationary velocity distribution, contrary to the
NFW one. The region of the cusp ($r<a$) is of main concern to us.

Since the gravitational potential $\phi(r)$ is constant, the specific energy $\ep=\phi(r)+v^2/2$
and the specific angular momentum $\vec K$ of each particle should be conserved. Instead of $\ep$,
it will be more convenient to use the apocenter distance of the particle $r_0$ (i.e. the maximum
distance on which the particle can move off the center, which can be found from the implicit
equation $\ep=\phi(r_0)+K^2/2r_0$). Being an implicit function of the integrals of motion $\ep$ and
$K$, $r_0$ is an integral of motion as well. Thus, any time variation of $\ep$, $\vec K$, or $r_0$
is necessarily a numerical effect, and we may judge the simulation convergency following the
behavior of these quantities.

We need to clarify two important points of our work. Some properties of the perfectly symmetrical
model we consider (like the exact conservation of the angular momentum for every particle) are
unstable and not realistic for real astrophysical DM halos that are always triaxial as a result of
tidal perturbations etc. The application of perfectly spherical models to real systems may give
rise to false conclusions \citep{pontzen2015}. However, the use of the spherical model for our
purposes is well founded. We are not considering the task of comparison of simulation results with
observations. Our aim is just to check if the 'N-body matter' behaves as a collisionless matter,
which is the principle question of the dark matter modelling.

Second, there is a frequent belief that it is much easier to converge on the spherically averaged
density distribution than on the full properties of the phase space distribution function. Indeed,
we need not correctly reproduce each individual particle trajectory. Moreover, it is not even
necessarily desirable since real dark matter halos are not spherically symmetric and therefore host
chaotic orbits. However, it would be completely wrong to disregard the phase evolution of the
system or consider its evolution as a 'second order effect' with respect to the density profile
shape.  As we will show in the \emph{Results: the simulation convergency} section, correct
simulations of the energy and angular momentum of each particle (contrary to individual particle
trajectories) are of critical importance for correct simulations of the density profile.

\begin{figure}
 \resizebox{\hsize}{!}{\includegraphics[angle=0]{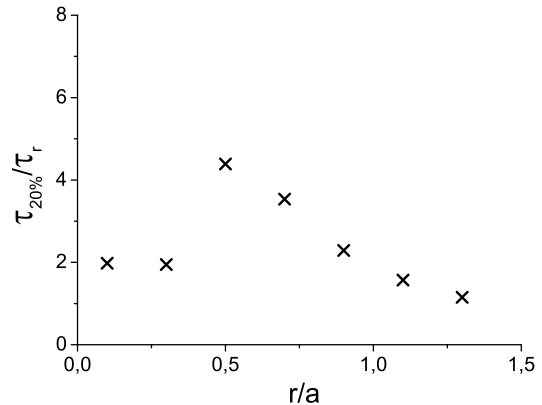}}
\caption{The ratio between the time $t_{20\%}$ when the mass inside some radius $r$ drops on more
than $20\%$, to the relaxation time $\tau_r(r)$.}
 \label{17fig1}
\end{figure}

\subsection{The simulations}
We simulate a single separate Hernquist halo. The aim of this work is to perform a sophisticated
test of the standard convergency criteria, therefore we do not try to model any real astronomical
object. Since the standard N-body units \citep{nbodyunits} are used, the results are independent on
the choice of $a$ and halo mass. However, we choose some values of the parameters, for illustrative
purposes. Let us set $a=100$~{pc}, which roughly corresponds to the well-known dwarf spheroidal
satellite of the Milky Way, Segue~1. This is one of the most popular objects for the indirect dark
matter search, since it is close to the Solar System; its present-day mass can be estimated as
$3\cdot 10^7 M_\odot$ \citep{12}. Segue~1 experienced strong tidal disruption, and we do not know
its initial mass. We consider two limiting cases. In the body of the paper we accept the halo mass
$M=10^9 M_\odot$, which is comparable with the present-day mass of a larger dwarf satellite, Fornax
\citep{walker2011} and almost certainly exceeds the initial mass of Segue~1. Thus we consider the
case of a compact and very dense dwarf spheroidal galaxy. However, since all the simulations are
performed in the dimensionless N-body units, a reader may easily extend the results for any value
of $M$. If $a$ is fixed, the only value that is sensitive to the choice of $M$ is time: all the
time intervals scale as $\Delta t\propto M^{-0.5}$ (while the ratios of time intervals remain the
same). As an illustration, we also considered the case of $M=10^7 M_\odot$, which is certainly
lower, then the present-day mass of Segue~1. The only difference is that all the time intervals get
ten times larger, and we everywhere specify the values corresponding to $M=10^7 M_\odot$ in the
footnotes. Anticipating events, we say that the results shown in all the figures in this paper are
not sensitive at all to the choice of $M$.

We use $N=10^6$ test bodies \footnote{All the data, as well as results of simulations of a Plummer
sphere of mass $10^{12}M_\odot$ we used as an auxiliary test model, are publicly available at {\it
http://www.das.uchile.cl/anton}}. They are placed randomly, in accordance with the analytically
obtained space and velocity distributions \citep{hernquist1990}. The relaxation time at $r=a$ is
$\tau_r(a)\simeq 8.8\cdot 10^{16}\text{s}\simeq 2.8\cdot 10^9$~{years}. Therefore, we make $200$
snapshots with the time interval $\Delta t=10^{15}\text{s}\simeq 30$~{mln. years}, covering the
time from $0$ to $t_{max}=2\cdot 10^{17}\text{s}\simeq 6.5\cdot 10^9$~{years} (for the case of the
halo mass $M=10^7 M_\odot$, $\tau_r(a)\simeq 2.8\cdot 10^{10}$~{years}, $\Delta
t=10^{16}\text{s}\simeq 300$~{mln. years}, $t_{max}=2\cdot 10^{17}\text{s}\simeq 6.5\cdot
10^{10}$~{years}). We record the positions and the velocities of each particle on each snapshot.

We evolve the system using one of the most extensively employed in cosmological simulations SPH
codes, Gadget-3, an update version of Gadget-2 \citep{springel2001,springel2005}. The gravitational
interactions in Gadget-3 are computed using a hierarchical tree \citep{tree1,tree2}. In this
algorithm the space is divided in different cells and the gravitational force acting on a particle
is computed using a direct summation for particles that are in the same cell and by means of
multipole (up to the quadrupole) expansion for the particles that are in a different cell. The
minimum distance between particles to be part of a different cell is controlled by a tree opening
criterion. Gadget-3 uses the Barnes-Hut tree opening criterion for the first force computation.
This criterion is controlled by an opening angle $\mu$, which determines the maximum ratio between
the distance to the center of mass of the cell ($d$) and the size of the cell ($l$). If the cell is
too close to the particle, $d/l$ will be greater than $\mu$, and new cells have to be opened to
maintain the accuracy on the force computation. In the further evolution of the system a dynamical
updating criterion (controlled by the fractional error $f_{acc}$) is used. We use the standard set
of parameters $\mu=0.7$ and $f_{acc}=0.005$, as suggested by \citep{springel2005, springel2008,
boylan-kolchin2009}. These values lead to a relative force error that is roughly constant in the
simulation $\sim 0.5$\%. We chose the softening radius $0.02a=2$~{pc}, in accordance with
\citep{kampen2000, hayashi2003}.

\begin{figure}
 \resizebox{\hsize}{!}{\includegraphics[angle=0]{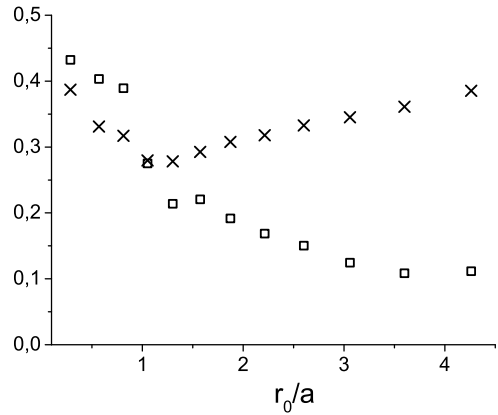}}
\caption{The averaged relative variations $\langle\widehat{\Delta} K/K_{circ}\rangle$ (squares) and
$\langle \widehat{\Delta} r_0/r_0\rangle$ (crosses) of the integrals of motion in a single time
step $\Delta t$. See the definition of the averaging $\widehat{\Delta}$ in the section {\it
Results: the integrals of motion}.}
 \label{17fig2}
\end{figure}

\section{Results: the integrals of motion}
Initially we convert each of the $201$ snapshots into the center-of-mass frame of references at the
moment when a snapshot is made.

First of all, we try to reproduce the results of \citep{power2003}. The density profile indeed
remains quite stable, and then a core in the center appears. Exactly following \citep{power2003},
we consider the moment $t_{20\%}$ when the mass inside some radius drops on $20\%$ comparing to the
initial value as the moment of the core formation. The ratio of $t_{20\%}$ to the relaxation time
$\tau_r$ at the same radius $r$ is represented in Fig.~\ref{17fig1}. We see that our data by and
large confirm the results of \citep{power2003}, the core really appears at $t\simeq 2\tau_r$.

\begin{figure}
 \resizebox{\hsize}{!}{\includegraphics[angle=0]{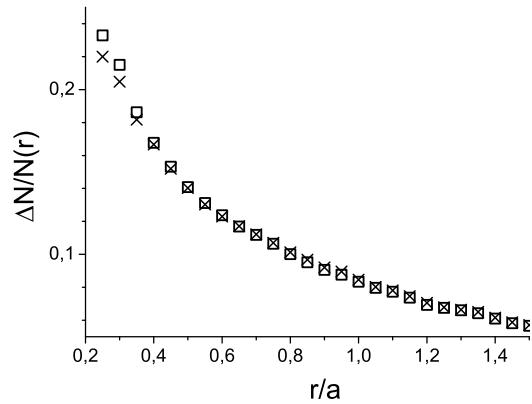}}
\caption{The upward $\Delta N_{+}(r)/\Delta t$ (squares) and downward $\Delta N_{-}(r)/\Delta t$
(crosses) Fokker-Planck streams of particles divided by the total number of the particles $N(r)$
inside $r$.}
 \label{17fig4}
\end{figure}

Before proceeding any further, two important comments relating to all the subsequent text should be
made. First, our convergency tests are mainly oriented on the radius interval $[0.25a;1.5a]$ where
they are the most precise. This choice of the working interval might appear strange at first sight:
typically the convergency problems occur much closer to the halo center. However, if we had chosen
a realistic area $(r\le 0.01a)$, then it would have contained only $\sim 100$ test particles, and
the statistic would have been poor. On the other hand, the density profile between $0.25a$ and
$0.75a$ remains much the same as in the center, since a power-law profile $\rho\propto r^{-1}$ is
self-similar. The lower border of the region under consideration $r=0.25a$ is defined by our choice
of the timestep $\Delta t=10^{15}$~{s} (for the case of the halo mass $M=10^7 M_\odot$, $\Delta
t=10^{16}\text{s}\simeq 300$~{mln. years}). At $r=0.1a$, $\tau_r\simeq\Delta t$, and the Hernquist
profile is certainly corrupted by the collisions even on the first timestep. However, as we will
see from the discussion of fig.~\ref{17fig4}, the core formation becomes visible in phase portrait
at much larger distances than in the density profile itself. Therefore, only the results related to
$r\ge 0.25a$ can be totally trusted.

Second, we want to consider variations of the integrals of motion as a function of radius. However,
each particle contributes to the density profile on an interval between its pericenter radius
$r_{min}$ and apocenter radius $r_0$. Hereafter we will consider $r_0$ as the characteristic radius
corresponding to the particle. Indeed, if the particle orbit is elongated, the particle spends
almost all the time near the apocenter, in accordance with the Kepler's second law. On the
contrary, if the orbit is circular, the particle moves along almost uniformly, but its radius
always remains close to $r_0$.

\begin{figure}
 \resizebox{\hsize}{!}{\includegraphics[angle=0]{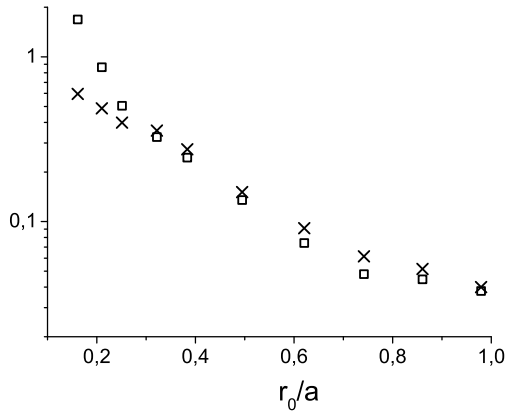}}
\caption{The ratios $\frac{K_{circ}}{\tau_r}\left\langle\frac{\widehat{\Delta} K}{\Delta
t}\right\rangle^{-1}$ (squares) and $\frac{1}{\tau_r}\left\langle\frac{\widehat{\Delta}
r_0}{r_0\Delta t}\right\rangle^{-1}$ (crosses) of the time, in which the particles completely
'forget' the initial values of the integrals of motion, to the relaxation time $\tau_r(r)$.}
 \label{17fig3}
\end{figure}

In order to study the behavior of the integrals of motion (theoretically they should conserve), we
order all $10^6$ particles according to their $r_0$ in the initial snapshot, and then divide the
particles into 200 groups of 5000 particles each. All the particles in the same group have similar
$r_0$, and the group may be characterized by the average initial $\overline{r_0}$ of its members.
We calculate $\Delta r_0/r_0=(r_0(i+1)-r_0(i))/r_0$ and $\Delta K/K_{circ}=(K(i+1)-K(i))/K_{circ}$
for each particle on each timestep. Here $i$ is the number of the snapshot, $K_{circ}$ is the
angular momentum corresponding to the circular orbit at $r_0$; apparently, this is the maximum
value of $K$ any particle with the apocenter distance $r_0$ may possess. Then we find the
root-mean-squares of $\Delta r_0/r_0$ and $\Delta K/K_{circ}$ averaged over each group and for each
snapshot. Our analysis shows that the root-mean-squares do not significantly depend on time until
the moment when the core forms at the radius corresponding to $\overline{r_0}$ of the group.
Therefore, we then average the root-mean-squares of $\Delta r_0/r_0$ and $\Delta K/K_{circ}$ over
all the timesteps where the core had not formed yet. We denote the values averaged in such a
complex manner by $\langle\widehat{\Delta} r_0/r_0\rangle$ and $\langle\widehat{\Delta}
K/K_{circ}\rangle$.

{The dependance of $\langle\widehat{\Delta} K/K_{circ}\rangle$ (squares) and
$\langle\widehat{\Delta} r_0/r_0\rangle$ (crosses) from the dimensionless radius $\overline{r_0}/a$
is represented in Fig.~\ref{17fig2}. We see that even in a single time step $\Delta
t=10^{15}\text{s}\simeq 30$~{mln. years} (for the case of the halo mass $M=10^7 M_\odot$, $\Delta
t=10^{16}\text{s}\simeq 300$~{mln. years}) the integrals (that should be constant) vary
significantly. Fig.~\ref{17fig3} represents the values
$\frac{K_{circ}}{\tau_r}\left\langle\frac{\widehat{\Delta} K}{\Delta t}\right\rangle^{-1}$
(squares) and $\frac{1}{\tau_r}\left\langle\frac{\widehat{\Delta} r_0}{r_0\Delta
t}\right\rangle^{-1}$ (crosses) that are the ratios of the time intervals in which an average
particle totally 'forgets' its initial values of $K$ and $r_0$ to the relaxation time $\tau_r(r)$.
Everywhere in the region of reliability ($r\ge 0.25a$) the ratios are much less than
$1$.\tolerance=20000

}It means that the particles totally 'forget' their integrals of motion in a time much shorter than
$\tau(r)$. In general one could not expect a reliable simulation of the velocity distribution at
$t\sim\tau(r)$ under such conditions.

\section{Results: the simulation convergency}
Thus, the integrals of motion of the particles are not conserved at all, while the density profile
remains stationary in quite good agreement with the theory in our simulations (we should mention,
however, that the absence of the collision influence up to almost $2$ relaxation times looks
surprising \citep{13}. The same stability and reproducibility of the cusps in cosmological
modelling leads to the wide acceptance of the idea that, though no significance can be attached to
the trajectories of individual particles in the N-body simulations, the cuspy density profile is
meaningful and should correctly describe the profiles of real halos. Let us use our results to
illustrate the vulnerability of the profile stability as the only convergency criterion of the
N-body simulations.

Indeed, if a Hernquist halo consists of real DM (we suppose that it is cold and noninteracting),
the values of $\ep$, $\vec K$, and $r_0$ of each particle must conserve, the particle distribution
function $f$ should depend only on $\ep$ and $K$ \citep{bt} and obey the collisionless kinetic
equation $df/dt=0$. It means that there are no particle fluxes in the phase space $(\ep,K)$.

However, figures~\ref{17fig2} and \ref{17fig3} doubtlessly reveal an intensive energy and angular
momentum exchange between the particles, i.e., the test bodies interact. Then the system may be
described by the Fokker-Planck (hereafter FP) equation \citep{ll10}
\begin{equation}
\dfrac{df}{dt}=\frac{\partial}{\partial q_\alpha}\left\{{\tilde A_\alpha}
f+\frac{\partial}{\partial q_\beta}[B_{\alpha\beta}f]\right\}
 \label{fokker_planck}
\end{equation}
where $q_\alpha$ is an arbitrary set of generalized coordinates,
\begin{equation}
{\tilde A_\alpha}=\dfrac{\overline{\delta q_\alpha}}{\delta t}\qquad
B_{\alpha\beta}=\dfrac{\overline{\delta q_\alpha\delta q_\beta}}{2\delta
t}
 \label{fokker_planck_coefficients}
\end{equation}
We may choose $q_1=\ep$, $q_2=K$, and figure~\ref{17fig2} shows that at least coefficients $B_{11}$
and $B_{22}$ in the equation (\ref{fokker_planck}) differ essentially from zero. Thus we model real
DM halos that are believed to be collisionless, by a system of test bodies governed by the kinetic
equation with a significant collisional term, i.e., by an {\it essentially collisional} equation.

\begin{figure}
 \resizebox{\hsize}{!}{\includegraphics[angle=0]{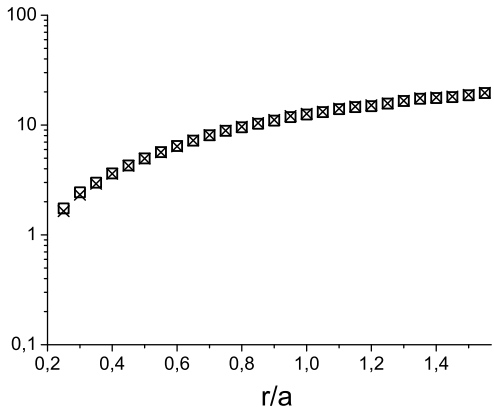}}
 \caption{The fractions of $N(r)$
particles inside radius $r$ that is carried away and in by the upward and downward Fokker-Planck
streams in the time $1.7\tau_r$ offered by \citep{power2003} ($1.7\tau_r\frac{\Delta
N_{+}(r)}{N(r)\Delta t}$ (squares) and $1.7\tau_r\frac{\Delta N_{-}(r)}{N(r)\Delta t}$ (crosses),
respectively).} \label{17fig5}\end{figure}

An important point must be underscored: the density profile in our simulation indeed holds its
shape (close to the NFW one in the center), in gratifying agreement with the theoretical
predictions. The variations of the integrals of motion, that we found, mainly touches on the
velocity distribution of the particles. Together with the UDP stability in cosmological
simulations, it can produce a dangerous illusion that N-body simulations might adequately model the
density profiles of dark matter structures, despite the fact that the velocity distribution was
distorted. We should emphasize that it cannot be true. Indeed, let us consider a stationary
spherically symmetric halo for the sake of simplicity. The particle distribution in the phase space
$f d^3x d^3v$ is a function of only the particle energy $\ep$ and three components if its angular
momentum $\vec K$ \citep{bt}. If the velocity distribution of the particles is anisotropic in each
point (which is the case under consideration in this paper), $f$ depends only on the particle
energy $f(ep)=f(\phi(r)+v^2/2)=f(\phi(r_0)+K^2/(2r_0^2))$. The particle speed distribution at some
radius $r$ is therewith equal to $4\pi v^2 f(\phi(r)+v^2/2) dv$, and the density is $\int 4\pi v^2
f(\phi(r)+v^2/2) dv$. These relationships clearly demonstrate the impossibility of a reliable
determination of the density profile without a reliable determination of the velocity distribution.
The distributions over $\vec v$ and $r$ are not just bound, there are actually a sort of
projections of the same distribution $f$ on the velocity or space coordinates. Apparently, this
conclusion is very general and does not depend on the assumption about the spherical symmetry that
we made.

We are able to compare the results with the theoretical prediction and check their agreement in the
model case that we consider. However, it is impossible in the case of real cosmological
simulations. Therefore, any numerical effect influencing on the velocity distribution $f(\vec v)$
or on the integrals of motion of the test particles puts the density profiles obtained in the
simulations in doubt. Moreover, the fact that the cuspy profile $\rho\propto r^{-1}$ turns out to
be very stable in our simulation, despite of $r_0$ and $K$ variations, is probably not a
coincidence, but a direct result of the numerical effects we discuss.

Indeed, the fact that we model collisionless systems with the test bodies governed by an
essentially collisional equation is surprising {\it per se}, but the main consequence is that the
profile stability does not guarantee the simulation correctness. The FP streams in the phase space
created by the particle interaction may form stable density profiles (corresponding to the
stationary solutions of the Fokker-Planck equation), but these profiles and their persistence are
at odds with the behavior of real collisionless systems. As the first and crude illustration, the
collisions lead to the contraction of the central region of any realistic profile and finally to
the core collapse. The density profile outside the core approaches a power law $\rho\propto
r^{-2.23}$ and then remains quite stable for a long time \citep{bt}. Of course, this distribution
is already formed by the unphysical test body collisions, and the immutability of the $\rho\propto
r^{-2.23}$ profile says nothing either about the simulation convergency or about the behavior of
real collisionless systems.

The core collapse appears at $t\gg\tau_r$ and has nothing to do with the Hernquist or the UDP
profiles. However, the Fokker-Planck equation has an another stationary solution close to the NFW
one \citep{evans1997,13}.

A question appears: if we obtain a stable cuspy density profile, how can we differentiate cusps
correlating with the properties of real collisionless systems from the solutions created by the
numerical effects?  In a collisionless system, the values of $r_0$ of the particles in the cusp
should remain constant. If collisions are significant, the values of $r_0$ should experience a
random walking, and the particles move up and down in the cusp forming a downward stream (of the
particles with decreasing $r_0$) and an upward stream (of the particles with increasing $r_0$). For
the cusp to be stable, the streams should compensate each other, which corresponds to a stationary
solution of the Fokker-Planck equation. Thus, if the cusp is created by the FP diffusion, we should
see two significant streams of particles with decreasing and increasing $r_0$, and the streams
should compensate each other in order to provide the cusp stability.

We chose two adjacent (i.e., divided by a single $\Delta t$) snapshots at the beginning of the
simulations, in order to minimize the core formation effects. For an array of radii $r$, we
calculated the number $\Delta N_{+}(r)$ of particles that had $r_0<r$ at the first snapshot and
$r_0>r$ at the second one, and the number $\Delta N_{-}(r)$ of particles that had $r_0>r$ at the
first snapshot and $r_0<r$ at the second one. Of course, $\Delta N_{+}(r)=\Delta N_{-}(r)=0$ in the
collisionless case, since $r_0$ is an integral of motion.

Fig.~\ref{17fig4} represents $\Delta N_{+}(r)$ (squares) and $\Delta N_{-}(r)$ (crosses) divided by
the total number of the particles $N(r)$ inside $r$. As we can see, the FP streams exist, though
they compensate well each other outside of $r=0.4a$. As we approach the center, the upward stream
becomes increasingly stronger than the downward one. This is the first sign of the core formation,
that is still invisible in the density profile at this radius, being already quite clear at the
phase picture of the system.

A question appears: are the discovered fluxes $\Delta N_{+}(r)$ and $\Delta N_{-}(r)$ real and
important? May they be just a small noise, produced by particles near the boundary, crossing and
recrossing it and thus giving the impression of flows that do not exist? One can readily see that
this is not the case. First of all, $\Delta N_{+}(r)$ and $\Delta N_{-}(r)$ apparently give only
the lower bounds on the upward and downward FP streams: the value of $r_0$ of a particle could have
crossed $r$ an odd number of times (and then it is counted only once) or an even number of times
(and then it is not counted at all). Since we count each particle no more than once on a timestep,
we totaly avoid the recrossing effect.

Second, the flows are just too strong to be just a noise. For instance, Fig.~\ref{17fig4} shows
that, though $\Delta N_{+}(r)$ and $\Delta N_{-}(r)$ are just the lower bounds on the streams,
$\Delta N_{+}(r)\simeq\Delta N_{-}(r)\simeq 2\cdot 10^4$ at $r=a$, i.e. $\sim 2 \%$ of the total
halo mass crosses this radius because of this unphysical effect on each timestep. This is
approximately the total number of particles in the layer of thickness $\sim a/12$ around the radius
$r=a$. The value $a/12$ by far exceeds the smoothing radius or any reasonable numerical noise that
may occur in the computing scheme.

The surprisingly high intensity of the Fokker-Planck diffusion is the main result of this work.
Approximately $8\%$ of particles are renewed even inside $r=a$. It means that in only $10\Delta
t\simeq 300$~{mln. years} (for the case of the halo mass $M=10^7 M_\odot$, $10\Delta t\simeq 3\cdot
10^{9}$~{years}), i.e., in $~5\%$ of the simulation time, all the particles inside the sphere $r=a$
(which contains a quarter of the total mass of the system) can be substituted by a purely numerical
effect. The fractions of particles inside radius $r$ that can be carried away or in by the upward
and downward Fokker-Planck streams in the Power's time $1.7\tau_r$ are $1.7\tau_r\frac{\Delta
N_{+}(r)}{N(r)\Delta t}$ and $1.7\tau_r\frac{\Delta N_{-}(r)}{N(r)\Delta t}$. Fig.~\ref{17fig5}
shows that they always significantly exceed $1$. The cusp in our simulations was created in the
initial conditions, but its shape is similar to the UDP. We can see that the unphysical FP streams
are strong enough to arbitrarily change the shape of the cusp (and therefore the shape is defined
by the FP diffusion rather than by the properties of the collisionless system) and even to create
it.

Another argument in support of the numerical nature of the cusps in cosmological simulations is the
profile universality. The similarity contradicts the observational results \citep{diversity}, but
is quite natural if the cusps are formed by the FP diffusion. An UDP-like stationary solution is
innate for the FP equation: the suppositions in \citep{evans1997} and \citep{13} are quite
different, but the results are similar. The properties of the solution of the FP equation are
totally defined by only a few coefficients $\tilde A_\alpha$ and $B_{\alpha\beta}$ that can be
similar for different N-body codes using similar algorithms, and are almost certainly the same
within the same simulation. As a result, the resulting halos are also self-similar, while nature is
much more variable.

The second important conclusion of this section is that the profile stability cannot be used as the
simulation convergency criterion: the first unquestionable signs of the influence of the test
particle interaction appear in the phase portrait much earlier than the density profile evolution
and the beginning of the core formation.

The third conclusion is that, since the variations of $K$ and $r_0$ are very significant in
Fig.~\ref{17fig2} even at $r>4a$, where the role of collisions or potential softening is minor, the
integral variations there are most likely due to the potential calculating algorithm. But whatever
the reason of the variations may be, the ill effect on the simulations is the same from the point
of view of the kinetic equation: variations if the integrals of motion reveal the collisional
influence and suggest that the system behavior is no longer described by the correct collisionless
equation. A convergence study with varying the opening angle $\mu$ of the Barnes-Hut tree opening
criterion, softening scale, as well as other parameters of the gravitational force computation, is
essential to understanding the origin of the non-conservation of integrals of motion and find the
optimal parameter set to decrease these undesirable numerical effects.

A much better understanding of the N-body simulation convergency is necessary to cast doubts on the
CDM model on the basis of 'cusp vs. core' contradiction.

\section{Acknowledgements}
The work is supported by the CONICYT Anillo project ACT-1122 and the Center of
Excellence in Astrophysics and Associated Technologies CATA (PFB06). We used the HPC clusters
Docorozco (FONDECYT1130458) and Geryon(2) (PFB06, QUIMAL130008 and Fondequip AIC-57).


\end{document}